\documentclass[twocolumn,showpacs,preprintnumbers,amsmath,amssymb,prl]{revtex4}

\usepackage{graphicx}
\usepackage{dcolumn}
\usepackage{bm}

\begin{document}

\preprint{APS/123-QED}

\title{Fast Quantum State Control of a Single Trapped Neutral Atom}

\author{M. P. A. Jones}
\author{J. Beugnon}
\author{A. Ga{\"e}tan}
\author{J. Zhang}
\altaffiliation[Now at ]{Key Laboratory of Quantum Optics, Ministry of Education - PRC and Institute of Opto-electronics, Shanxi University, China}
\author{G. Messin}
\author{A. Browaeys}
\author{P. Grangier}
\affiliation{Laboratoire Charles Fabry de l'Institut d'Optique (UMR 8501), Campus Polytechnique, RD 128, 91127 Palaiseau Cedex, France}

\date{\today}

\begin{abstract} We demonstrate the initialisation, read-out and high-speed manipulation of a qubit stored in a single
 $^{87}$Rb atom trapped in a submicron-size optical tweezer. Single qubit rotations are performed on a
sub-100 ns time scale using two-photon Raman transitions. Using the ``spin-echo'' technique, we measure an irreversible
dephasing time of 34 ms. The read-out of the single atom qubit is at the quantum projection noise limit when averaging up to 1000 individual events.
\end{abstract}

\maketitle
 The building block of a quantum computer is a qubit~-~an isolated two-level quantum system on which
one can perform arbitrary single-qubit unitary operations. In the circuit approach to quantum computing~\cite{nielsen},
 single qubit operations are sequentially combined with two-qubit gates to generate entanglement and
realise arbitrary quantum logic operations. In the alternative one-way quantum computing scheme~\cite{raussendorf01}, the ensemble of
qubits is prepared in a highly entangled cluster state, and computations are performed using single qubit
operations and measurements. A wide range of physical systems are under investigation as potential qubits
\cite{nielsen,leggett,leibfried03}, including trapped single neutral atoms. In particular, the hyperfine ground states of
alkali metal atoms can be used to make qubits that are readily manipulated using microwave radiation or Raman
transitions, with
negligible decoherence from spontaneous emission. The usefulness of these techniques has been demonstrated with the
realisation of a 5-qubit quantum register based on microwave addressing of single atoms trapped in an optical
lattice~\cite{schrader04}. A quantum register could also be formed using arrays of optical tweezers \cite{dumke02}, each
containing a single atom~\cite{bergamini04}, with each site optically addressed using tightly focussed Raman
beams~\cite{saffman06}. Several detailed proposals for performing two - qubit operations in such a tweezer array have been made, based on controlled collisions~\cite{Dorner05}, dipole-dipole interactions between Rydberg atoms~\cite{protsenko02,Saffman05} and cavity-mediated photon
exchange~\cite{you03}.
Alternatively, two-qubit operations could be performed without interactions by using photon
emission~\cite{darquie05} and quantum interference effects~\cite{Duan06}. The recent observations of atom-photon entanglement~\cite{blinov04,volz06} and two-photon interference between single photons emitted by a pair of trapped atoms~\cite{beugnon06,maunz06} are major steps in this direction.

\begin{figure}
\includegraphics{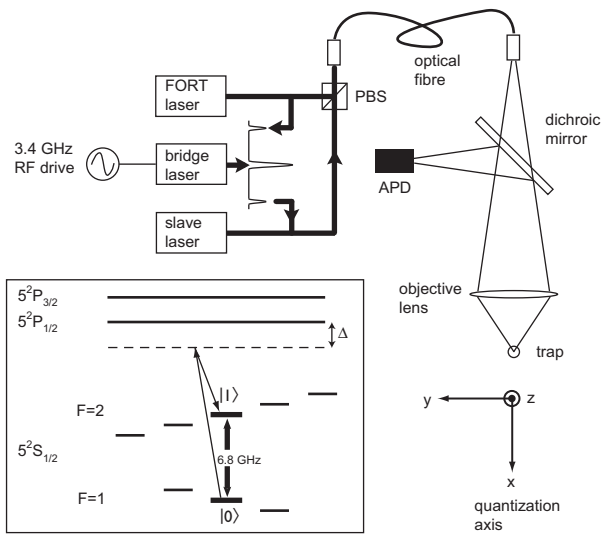}
\caption{\label{setup}Experimental setup. A high performance objective lens creates a tightly focussed optical dipole
trap, which also acts as one of the Raman beams. The second Raman beam is generated using two additional
 diode
lasers, and is superimposed with the trapping beam on a polarising beam splitter (PBS). A single polarization-maintaining
fibre carries both beams to the experiment. Inset shows the relevant energy levels of $^{87}$Rb. The quantisation axis is
defined by a 0.36 mT magnetic field along the $x$-axis.}
\end{figure}

In this paper we describe how a single $^{87}\mathrm{Rb}$ atom trapped in an optical tweezer can be used to store,
manipulate and measure a quantum bit. The qubit basis states are the $|0\rangle = |{F=1,m_{F}=0}\rangle$ and
$|1\rangle=|{F=2,m_{F}=0}\rangle$ ground state hyperfine sublevels (Fig.~\ref{setup}). We initialise the system by preparing
the atom in the $|0\rangle$ state using optical pumping. Single qubit operations are performed using two-photon Raman
transitions.  A novel feature of our experiment is that we use the tightly focussed optical tweezer as one of the Raman
beams. In this way we obtain a Rabi frequency of $\Omega =2 \pi \times 6.7$ MHz with laser beams detuned by $>10^6$
linewidths from the nearest atomic transition. We perform a measurement of the
state  ($|0\rangle$ or $|1\rangle$) of each single atom with near unit efficiency, allowing us to perform projection-noise
limited measurements of the qubit state. Using Ramsey spectroscopy, we measure a dephasing time of 
$370\ \mu$s. This dephasing can be
reversed using the spin-echo technique. In this way we have measured an irreversible dephasing time 
of $34$ ms, which is almost six orders of magnitude longer than the time required to perform a $\pi$ rotation. Due to
the very large Rabi frequency,  this ratio can approach the state of the art achieved in ion trap systems that have much
longer coherence times \cite{langer05}.

We isolate and trap single $^{87}\mathrm{Rb}$ atoms in an optical dipole trap created by a tightly focussed far-off
resonant laser beam \cite{schlosser01}. A custom-made objective lens with a numerical aperture of 0.7 is used to focus
the beam
at 810~nm to a diffraction-limited waist of $\approx  0.9\ \mu$m. With a power of 0.95\ mW we obtain a trap at the focus
with
a depth of 1.2~mK and oscillation frequencies of 125 kHz and 23 kHz in the radial and axial directions respectively.
The trap is loaded from an optical molasses. A ``collisional blockade'' effect forces the number of atoms in the trap to be either zero or
one. We measure the initial temperature of the atom to be  $90~\mu$K. The trap lifetime  in
the absence of any near-resonant light is 3~s and the heating rate is
$0.021\pm0.005\ \mu\mathrm{K}\ \mathrm{ms}^{-1}$.

The presence of an atom in the trap is detected using its fluorescence from the molasses cooling light.
As shown in Fig. \ref{setup}, the fluorescence is collected by the same objective used to make the optical dipole trap,
and is separated off using
a dichroic mirror before being imaged on to an avalanche photodiode. When an atom is present in the trap we detect
 $\approx10,000$ photons s$^{-1}$, compared to a typical background count rate of 2000 s$^{-1}$ for an empty
trap. By
setting a threshold value for the fluorescence we can unambiguously detect the presence of an atom within $\approx 15$~ms of its arrival. This signal is then used to shut off the molasses light and trigger the
experimental sequence. 

Once a single atom has been detected in the trap, it is prepared in the logical state $|0\rangle$ using optical pumping on the $D2$ line.
For this we use a $\pi$-polarized Zeeman pumping beam resonant with the $F=1 \rightarrow F^{\prime}=1$ transition and a hyperfine repumping beam
 resonant with the $F=2 \rightarrow F^\prime=2$
transition. The quantization axis is defined by a 0.36~mT magnetic field along the $x$-axis. After 200 $\mu$s of
optical pumping the atom is prepared in the logical state $|0\rangle$ with 85\% efficiency. We have determined that
all of the atoms not prepared in  $|0\rangle$ are left in the other $F=1$ sublevels. These atoms are not affected by the
Raman beams due to the Zeeman shift.

We perform single qubit rotations by coupling the logical states $|0\rangle$ and $|1\rangle$ using a two-photon stimulated
Raman transition. Driving the Raman transition requires two phase-locked laser beams separated by the
hyperfine transition frequency $\omega_{\mathrm{hf}} /2\pi \simeq  6.8$ GHz. In our experiment, the optical dipole trap
forms one of these beams. The trapping light is produced using a grating stabilised external cavity diode laser. 
To generate the second Raman beam we use two additional 810~nm diode lasers as shown in Fig.
\ref{setup}. The frequency offset is obtained by modulating the current of the bridge laser at 3.4 GHz, adding
two sidebands to its output with the desired frequency separation. The bridge laser is phase-locked to the dipole trap
laser by injection locking
on one of the sidebands \cite{snadden97}. A Mach-Zender interferometer is used to remove 90 \% of the carrier power from
the output of
the bridge laser,  which is then used to injection lock a third slave laser tuned to the other sideband. An acousto-optic modulator allows 
intensity control of the Raman beam as well as fine tuning of the frequency difference between the two beams. The two
beams are sent to the experiment through the same polarization-maintaining optical fibre. The optical dipole trap and
the Raman beam have orthogonal linear polarizations in the $z-y$ plane in order to drive $\Delta m_{F}=0$ transitions.

After the Raman beams have been applied, we measure the state ($|0\rangle$ or $|1\rangle$) of
the atom. A probe
laser beam resonant with the $5^{2}S_{1/2}\,F=2 \rightarrow 5^{2}P_{3/2}\, F^{\prime}=3$ cycling transition is used to
state-selectively push atoms in
 state $|1\rangle$ out of the trap by radiation pressure. During the 100 $\mu$s that the probe beam is applied, the depth
of the trap is lowered to 0.4 mK to make sure that atoms in $|1\rangle$ are rapidly removed from the trap before they can
be pumped into the $F=1$ hyperfine level by off-resonant excitation. Atoms that are initially in state $|0\rangle$ are
unaffected by this procedure and remain in the trap \cite{atomloss}. We then turn on the molasses cooling light for
10 ms and determine whether or not the
atom is still in the trap. The states $|0\rangle$ and $|1\rangle$ are therefore mapped onto the presence (absence) of the
atom at the end of
the sequence, as was shown in similar experiments with caesium atoms
\cite{schrader04,kuhr05}. 

This technique actually measures whether the atom is in the $F=1$ or $F=2$ hyperfine level at the end of the sequence.
Therefore, atoms that are left in the ${F=1,m_{F}\pm1}$ sublevels after optical pumping also contribute to the signal,
leading to a 15 \% background on the probability that the atom remains after the push-out laser is applied. 
To independently check how accurately we can determine whether an atom is in the $F=1$ or $F=2$ hyperfine state, we prepare the atom in either ${F=1}$ or ${F=2}$ by blocking one of the optical pumping beams. We measure that
the probability that we have incorrectly assigned the hyperfine level of the atom at the end of a single sequence is
less than 2\%.

At the end of a single qubit operation, the qubit is in general in a superposition $ \alpha |0\rangle + \beta
|1\rangle$. In order to measure the coefficients $\alpha$ and $\beta$ we repeat each experiment (trapping, preparation,
qubit operation and readout) 100 times under identical conditions. In the absence of technical noise, the statistical
error on the mean recapture probability after $N$ identical experiments should be given by the standard deviation of
the binomial distribution $\sigma = \sqrt{p (1-p)/N}$ where $p$ is the probability that the atom is in $F=1$. We have
checked experimentally that this is the case for values of $p$ between 0.005 and 0.95, and for $N$ up to 1000. Our
measurements of the coefficients $\alpha$ and $\beta$ are therefore limited solely by quantum projection noise.

\begin{figure}
\includegraphics{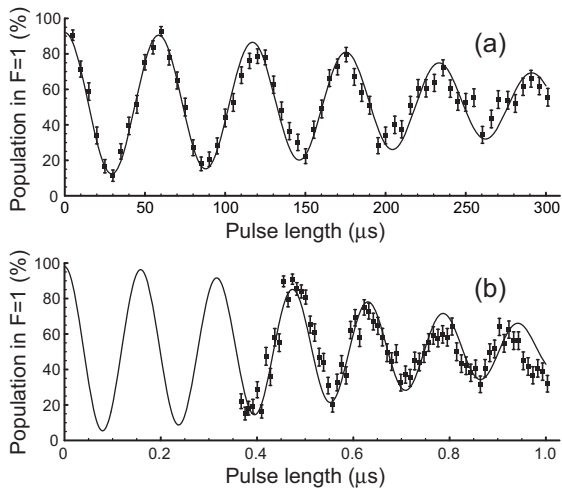}
\caption{\label{rabi}Single-atom Rabi oscillations. We measure the fraction of atoms in $F=1$ as a function of the Raman
pulse length, at low (a) and high (b) intensity. We observe damped Rabi oscillations between the two qubit states with
Rabi frequencies of $\Omega = 2\pi \times 18$ kHz (a) and $\Omega = 2\pi \times 6.7$ MHz (b). In (b) we could not
observe the first 400~ns due to the response time of the acousto-optic modulator. The error bars correspond to
the quantum projection noise.}
\end{figure}
The combined performance of these techniques was investigated by performing Rabi rotations between the states $|0\rangle$
and $|1\rangle$. The results for two different Raman beam intensities are shown in Fig. \ref{rabi}. At our maximum
intensity, we reach a Rabi frequency of $\Omega = 2\pi \times 6.7$ MHz, which corresponds to a $\pi/2$ rotation time of
37~ns. The 15 \%  background is due to the imperfect optical pumping discussed above. At both high and low power the
oscillations are strongly damped, decaying after approximately 5 complete periods. We attribute this damping to
intensity fluctations in the Raman beams, due both to technical intensity noise (we measure $\approx 2$ \% RMS on each beam), and to the time varying intensity seen by the atom due to its motion. The latter  is modelled by averaging the Rabi frequency over the motion of the atom, assumed to be thermal \cite{kuhr05}. The solid lines in Fig. \ref{rabi}(a) and (b) are fits using
a model which includes both effects. For both curves the temperature is fixed at  $90\ \mu$K and the 
 total technical intensity noise  (both beams) is fixed at 2.5 \%. The initial contrast and the Rabi
frequency are the only adjustable parameters. The model is in 
good agreement for both curves, despite the $130,000$ fold reduction in the Raman beam intensity (using neutral
density filters) between the two curves

\begin{figure}
\includegraphics{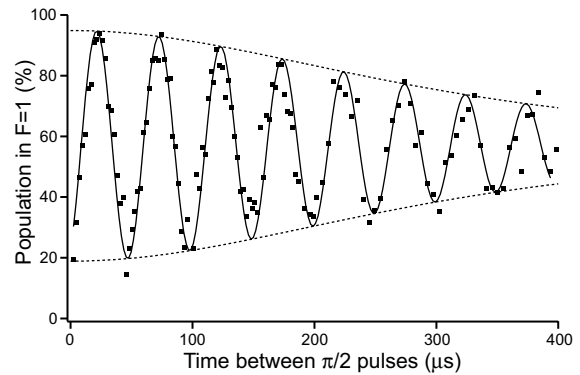}
\caption{\label{ramsey} Ramsey fringes recorded with a $\pi/2$ pulse length of $1.2\ \mu$s and a detuning $\delta = 2\pi
\times 20.8$~kHz. The solid line is a fit  using the model presented in \cite{kuhr05}, which yields a
dephasing time $T_{2}^{*}=370\ \mu$s. The dotted line is the envelope of this fit.}
\end{figure}
We have also investigated the coherence properties of this qubit using Ramsey spectroscopy. We apply two $\pi/2$ pulses
separated by a variable time $t$, with a fixed value of the Raman detuning $\delta$. In the limit $\delta \tau \ll
1$ where $\tau$ is the $\pi/2$ pulse length, the population measured in the $|1\rangle$ state varies as $ P(t)=\mathrm{cos}^2(\delta
t/2)$. The results of this measurement with $\tau=1.2\ \mu$s and $\delta = 2\pi \times 20.8$ kHz are shown in Fig.
\ref{ramsey}. The
contrast of the interference fringes decays as the time between the two $\pi/2$ pulses is increased, with a $1/e$ decay
time of approximately $370\ \mu$s due to dephasing of the atomic qubit compared to the Raman beams. The
dephasing mechanisms that operate in optical dipole traps have been extensively studied
\cite{kuhr05}.
In our case, the dominant dephasing mechanism arises from the finite temperature of the atoms in the trap. Due to the
6.8~GHz hyperfine splitting, the detuning of the dipole trap $\Delta$ is slightly different for the $|0\rangle$ and
$|1\rangle$ states, which therefore experience slightly different AC Stark shifts. This gives rise to a position
dependence of the qubit transition frequency $\omega(\bm{r})=\omega_{\mathrm{hf}}+\eta U(\bm{r})/\hbar$, 
where the differential AC Stark shift coefficient $\eta\ (\approx \omega_{\mathrm{hf}}/\Delta) = 7 \times 10^{-4}$ for
our trap. Averaged over the motion of the atom in the trap, this effect shifts the detuning $\delta$ between the atomic
resonance and the Raman beams by an amount which is different for each atom in a thermal ensemble, depending on its
energy. As shown in \cite{kuhr05}, this gives rise to a decay of the contrast with a characteristic (1/e) decay time
$T_{2}^{*} = 1.94 \hbar/\eta k_{B} T$. We measure a dephasing time of $T_{2}^{*}=370\ \mu$s, which is longer
than the theoretical value  $T_{2}^{*}=220\ \mu$s that we would expect at $90\ \mu$K. By
varying the temperature we have confirmed that the dephasing is due to the motion of the atom, although this 
quantitative disagreement  remains unexplained. 
\begin{figure}
\includegraphics{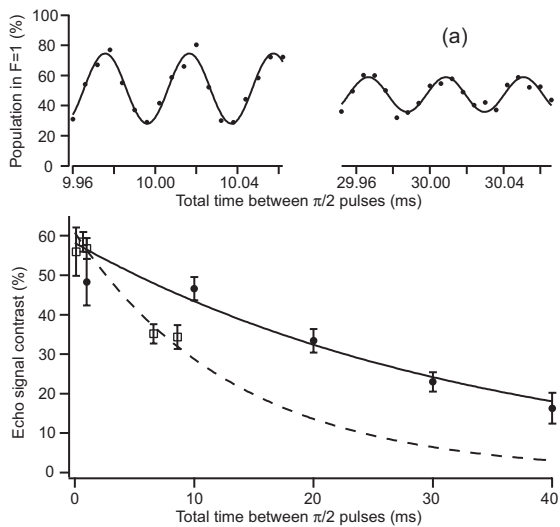}
\caption{\label{echo}(a) Example of the echo signal. We fix the time between the first $\pi/2$ pulse and the $\pi$ pulse at $T=5$~ms (left) and $T=15$~ms (right), and vary the time of the second
$\pi/2$ pulse around $t=2T$. The trap depth is $U=0.4$~mK, and the magnetic field is $B=0.18$~mT. (b) Echo
signal contrast as a function of the total time between the $\pi/2$ pulses with
$U=1.2$~mK and $B=0.36$~mT (open squares) and $U=0.4$~mK and $B=0.18$~mT (filled circles). The dashed and solid lines are
exponential fits with $1/e$ decay times of $13\pm2$~ms  and $34\pm5$~ms respectively. }
\end{figure}

The dephasing due to the motion of the atoms in the trap can be reversed
using the ``spin-echo'' technique \cite{andersen03,kuhr05}. An
additional population-inverting $\pi$ pulse applied midway between the two $\pi/2 $ pulses ensures that the
phase accumulated during the second period of free evolution is the opposite of that acquired during the first. The
echo signals that we obtain are shown in Fig. \ref{echo}. The echo signal decays due to the decay
of the populations ($T_{1}$ processes) and the loss of atoms to other Zeeman states, as well as irreversible dephasing
caused by fluctuations in the experimental parameters. 
To illustrate this, we repeated the spin echo experiments with a reduced trap depth and a smaller magnetic field.
Lowering the trap depth reduces the rate of hyperfine mixing due to spontaneous Raman transitions
induced by the optical dipole trap, and reducing the magnetic field reduces the sensitivity of the qubit states to
ambient
magnetic field fluctuations. As shown in Fig. \ref{echo}, this resulted in a significant increase in the decay time
of the echo signal from 13~ms to 34~ms.

In conclusion, we have demonstrated that a single rubidium atom trapped at the focal point of a large numerical aperture lens is a  promising system for encoding a qubit. With improved state preparation and the elimination of technical intensity noise, the fidelity of our single qubit operations will ultimately be limited by the motion of the atom. This effect could be reduced by further laser cooling. As well as their importance for high-speed quantum logic, fast  single qubit operations are also important in many entanglement schemes. Most existing protocols require atoms initialised in $|0\rangle$  to be rotated into a superposition state before the entanglement operation is applied~\cite{Saffman05,barrett05}. Several  proposals for generating entanglement using photon emission also require state rotation between successive photons~\cite{barrett05,beige05}. In our
experiment, we can generate single photons at a rate of 5 MHz~\cite{beugnon06}, leaving just 200\,ns between successive photon emission events in which to perform a qubit rotation. Here we show that we can perform single qubit  rotations on this timescale, and thus avoid limiting the rate at which entangled pairs can be created and gate operations  performed in our system.

\begin{acknowledgments}
This work was supported by Institut Francilien des Atomes Froids (IFRAF), ARDA/DTO and the Integrated Project SCALA
from the European IST/FET/QIPC program. M. P. A. Jones was supported by Marie Curie fellowship no. MEIF-CT-2004-009819. 
We thank the laser cooling group at NIST Gaithersburg, Poul Jessen, Ivan Deutsch and Gerhard Birkl for helpful discussions.
\end{acknowledgments}


\begin{thebibliography}{99}
\bibitem{atomloss}We observe a small (5\%) loss due to the reduction in trap depth. This reduces the initial
contrast in Fig.~2
\end{thebibliography}
\end{document}